\documentclass[preprint,onecolumn,nofootinbib]{revtex4}

\usepackage{mathrsfs} 
\usepackage{amssymb}
\usepackage{slashed} 
\usepackage[graphicx]{realboxes}  
\usepackage{tikz} 
\usepackage{subcaption} 
\usepackage{pgfplots} 
\usepackage{booktabs}  
\usepackage{comment}

\usepackage[colorlinks=true,linkcolor=blue,urlcolor=blue,filecolor=black,citecolor=red,pdfstartview=FitV,pdftitle={},pdfsubject={},pdfkeywords={},pdfpagemode=None,bookmarksopen=true]{hyperref}  
\usepackage{graphicx}
\usepackage{amsmath}
\usepackage{amsfonts}  
\usepackage{amssymb,ulem}
\usepackage{color,xcolor}%
\usepackage{CJK}
\usepackage{amsthm,amsmath,amssymb} 
\usepackage{mathrsfs}
\usepackage{url}
\usepackage{hyperref}
\usepackage{array}

\usepackage{dcolumn}
\usepackage{float}
\usepackage{appendix}
\newcommand{\red}{\textcolor[rgb]{1.00,0.00,0.00}}

\newcommand{\f}{\begin{equation}}
\newcommand{\ff}{\end{equation}}
\newcommand{\fa}{\begin{eqnarray}}
\newcommand{\ffa}{\end{eqnarray}}

\begin{document}

	\captionsetup{justification=raggedright}

\title{Quasinormal modes of a rotating loop quantum black hole}

\author{Zhongzhinan Dong$^{1}$}
\thanks{gs.zzndong25@gzu.edu.cn}
\author{Shulan Li$^{2}$}
\thanks{shulanli.yzu@gmail.com}
\author{Dan Zhang$^{3}$}
\thanks{danzhanglnk@163.com}
\author{Jian-Pin Wu$^{3}$}
\thanks{jianpinwu@yzu.edu.cn}

\affiliation{
$^{1}$ \mbox{School of Physics, Guizhou University, Guiyang 550025, China}\\
$^{2}$ \mbox{Department of Physics, Shanghai University, Shanghai 200444, China}\\
$^{3}$ \mbox{Center for Gravitation and Cosmology, College of Physical Science and Technology,} \mbox{Yangzhou University, Yangzhou 225009, China}
}

\begin{abstract}

We investigate the quasinormal modes of a massless scalar field on an effective rotating loop quantum black hole background, constructed from a covariant spherical model via an improved Newman-Janis algorithm. Using the continued fraction method, we compute the spectrum for both fundamental and overtone modes, and systematically analyze how the frequencies depend on the quantum correction, spin, and angular structure of the perturbation. For all fundamental modes, increasing the quantum gravity correction monotonically reduces both the oscillation frequency and the damping rate, signaling slower oscillations and prolonged decay. Rotation imprints a nontrivial modulation: for a spherically symmetric perturbation, the real frequency displays a crossover as the spin grows, whereas this feature is suppressed once angular momentum is turned on; further activating the azimuthal component enhances the frequency and reduces the damping even more strongly. In the overtone sector, the rotating solution retains the hallmark quantum gravitational signatures of the spherical case — overtone outbursts and non-monotonic evolution — with rotation shifting these phenomena to weaker quantum corrections. Nonzero orbital angular momentum suppresses the outbursts, while the azimuthal degree of freedom boosts the frequency, giving rise to novel spectral inversions among higher overtones. Our results confirm that the effective rotating metric captures essential loop quantum gravity features, providing clear theoretical benchmarks for black hole spectroscopy and future gravitational-wave observations.

\end{abstract}
	
	\maketitle
	\tableofcontents

	\section{Introduction}
	
The singularity problem, inherent in black holes (BHs) as predicted by general relativity (GR), represents a fundamental challenge in classical gravitational theory. The existence of singularities signifies a breakdown of spacetime continuity and the known laws of physics, revealing the limitations of classical theory under extreme conditions. To address this, researchers have pursued new avenues of inquiry. A complete theory of quantum gravity is now widely regarded as the most promising path toward a fundamental resolution to this problem \cite{Lan2023,Perez2017,Modesto2007}.

Among candidate theories of quantum gravity, loop quantum gravity (LQG) is particularly notable for its non-perturbative and background-independent framework. It has achieved substantial success in cosmology, where quantum geometric effects are shown to replace the classical Big Bang with a ``Big Bounce'', thereby resolving the initial singularity \cite{Ashtekar2006,Ashtekar2006a,Ashtekar2006b}. Building on this, LQG-inspired quantum corrections have also been applied to investigate the potential origins of dark energy and inflationary dynamics \cite{Bojowald2003,Ashtekar2021,Bojowald2007,Bojowald2002}.

Extending the symmetry reduction strategy from loop quantum cosmology (LQC) to BHs has led to a class of effective BH models with LQG corrections, primarily implemented via holonomy modifications \cite{Modesto2010,Modesto2006,Campiglia2007,Gambini2020,Boehmer2007,Chiou2008,Yang2023,Gan2024}. These models share a crucial feature: the classical singularity is superseded by a finite, regular transition surface bridging trapped and anti-trapped regions.
This structure not only resolves the curvature divergence but also encodes new physics arising from quantum geometry, offering a novel perspective on the internal structure and quantum aspects of BHs.

Existing studies have shown that, in the late-time ringdown phase of binary BH mergers, the gravitational waves (GWs) are predominantly governed by the BH's quasinormal modes (QNMs) \cite{Teukolsky1973,Echeverria1989,Finn1992}. These modes are characteristic damped oscillations triggered by perturbations, with their frequencies and decay rates uniquely determined by the BH's fundamental parameters. Consequently, the information encoded in GW signal can reveal the internal structure and geometric features of BHs, offering crucial insights into its stability and spacetime properties \cite{Berti2006, Berti2018, Gong2024, Zhu2025}.
	
Early analyses of gravitational-wave ringdown signals typically focused on the fundamental mode ($n=0$). In recent years, however, advances in detector sensitivity and data analysis techniques have brought the role of overtones ($n>0$) into sharp relief \cite{Isi2019,Abbott2021}. Studies indicate that during the early ringdown phase, overtones contribute more significantly than the fundamental mode. Consequently, relying solely on the latter fails to recover accurately the true mass and spin of the remnant BH, whereas including overtones substantially improves the precision of parameter estimation \cite{Giesler2019}. Further research has revealed a connection between overtone signatures and the geometry of the BH event horizon, showing that even subtle structural variations can strongly influence the first few overtones \cite{Konoplya2024}. These insights not only open new avenues for probing horizon geometry through overtone analysis but also motivate deeper investigations into overtone behavior across different BH spacetimes. Although practical challenges — such as noise in real observations — can interfere with and bias overtone parameter estimation \cite{Bhagwat2018,Cotesta2022}, the foundational importance of overtones in gravitational-wave ringdown studies remains clearly established \cite{Konoplya2022,Konoplya2023,Konoplya2023a}.

The unique correspondence between quasinormal frequencies (QNFs) and BH parameters implies that quantum gravitational effects should leave distinctive imprints on their QNF spectra. This provides a promising avenue for probing quantum gravity effects through gravitational-wave astronomy, making BH spectroscopy a key window into candidate quantum theories. 

While early spherically symmetric loop quantum black hole (LQBH) models have been successful in resolving classical singularities, they generally break full spacetime covariance, limiting the theoretical consistency of the quantum corrections. In contrast, covariant LQBH models preserve the underlying LQG constraint structure and maintain covariance, providing a robust and reliable framework for quantum gravity effects. Motivated by these advantages, a covariant LQBH characterized by the quantum parameter $r_0$ or $\bar{\lambda}$, commonly referred to as the ABBV BH \cite{AlonsoBardaji2022,AlonsoBardaji2022a}, has attracted considerable attention and stimulated systematic investigations of its QNM properties. In \cite{Fu2024}, the QNM spectrum of this model under massless scalar and electromagnetic field perturbations was first analyzed using the pseudo-spectral method. Subsequent studies extended the analysis to massive scalar, electromagnetic, and Dirac field perturbations \cite{Bolokhov:2023bwm}. These works revealed that the QNM spectra generically exhibit overtone outbursts and oscillatory behavior across different types of perturbations. Building upon these results, Alonso-Bardaji \textit{et al.} further generalized the model to include electric charge and a cosmological constant \cite{Alonso-Bardaji:2023niu}. More recently, \cite{Zhu:2024wic} investigated the QNM spectrum of scalar and Dirac fields in charged ABBV BH spacetimes, showing that while either the quantum parameter or the electric charge alone can trigger overtone outbursts, their combined effect tends to suppress such phenomena.

Given that most existing studies of LQBHs have been restricted to spherically symmetric configurations, it is important to note that typical astrophysical BHs are in fact rotating. However, constructing effective rotating LQBH models remains highly challenging due to technical obstacles associated with real-valued Ashtekar–Barbero variables in axisymmetric spacetimes \cite{Bojowald2006,Gambini2011,Gambini:2020fnd,Frodden:2012en}. As a result, a mathematically consistent effective model for rotating LQBHs is still lacking. 

One alternative route is to employ the modified Newman–Janis (NJ) algorithm \cite{AzregAinou2014}, where a static LQG metric is used as a seed to generate an effective rotating counterpart. While the applicability of the NJ algorithm beyond GR requires further scrutiny, the resulting models can still capture key features of loop-quantum-corrected BHs. These features not only provide observational signatures testable with gravitational-wave observation or the Event Horizon Telescope, but may also help constrain or rule out certain loop quantization prescriptions \cite{Liu:2020ola,Brahma:2020eos,Chen:2022nix}. Motivated by the above idea, this work starts from the ABBV BH, applies the NJ algorithm to obtain its rotating solution, and investigates the fundamental modes and overtones of scalar field perturbations.
	
This paper is structured as follows. In Sect.\ref{section2}, we briefly review the spherically symmetric LQBH that is chosen as the seed metric and construct its effective rotating solution using the NJ algorithm. In Sect.\ref{section4}, we use the continued fraction method to obtain the QNFs and analyze their characteristics. The conclusions and discussions are presented in Sect.\ref{section5}.

\section{An effective rotating loop quantum black hole}\label{section2}

In this section, we employ the NJ algorithm to construct an effective rotating BH solution within the framework of LQG. Our starting point is the ABBV BH proposed in Refs. \cite{AlonsoBardaji2022,AlonsoBardaji2022a}. Although a fully self-consistent effective rotating LQBH model derived from first principles remains an open challenge, the NJ algorithm offers a tractable and physically well-motivated approach to incorporating rotation. The resulting effective metric captures essential features expected from LQG. 

The exterior geometry of the ABBV BH \cite{AlonsoBardaji2022,AlonsoBardaji2022a} is described by the metric:
\begin{equation}\label{spherically symmetric black hole metric}
\begin{aligned}
		&ds^2=-f(r)dt^2+\frac{1}{g(r)f(r)}dr^2+r^2d\Omega^2\,,\\
		&f(r)=1-\frac{2M}{r},\qquad g(r)=1-\frac{r_0}{r}\,,
\end{aligned}
\end{equation}
 where $r_0~<~2M$ is a fundamental length scale introduced by LQG effects. This parameter is defined as:
\begin{equation}
r_0=2M\frac{\bar{\lambda}^2}{1+\bar{\lambda}^2}\,,
\end{equation}
in which $\bar{\lambda}$ is a dimensionless constant associated with the fiducial length of the holonomies and $M$ is a constant of motion. This length scale $r_0$ implies the existence of a minimum area proportional to $r_0^2$ in this model \cite{AlonsoBardaji2022,AlonsoBardaji2022a}. In the limit $\bar{\lambda}\rightarrow 0$, one finds $r_0\rightarrow 0$, and the effective LQG geometry described by Eq.~\eqref{spherically symmetric black hole metric} reduces to the classical Schwarzschild solution.

We now proceed to construct the rotating counterpart of the above spherically symmetric LQBH model using a modified NJ algorithm. Originally introduced by Newman and Janis in 1965 \cite{Newman1965}, the NJ algorithm has been widely applied to generate rotating BH metrics from static seeds. However, as noted in \cite{AzregAinou2014}, a key ambiguity in the traditional approach lies in the complexification of the radial coordinate $r$. Different choices of complexification can lead to distinct — and sometimes unphysical — solutions, which may not admit a representation in Boyer–Lindquist coordinates (BLCs). Moreover, due to this ambiguity, the final step of transforming from Eddington–Finkelstein coordinates (EFCs) to BLCs often fails. To overcome these issues, the authors of \cite{AzregAinou2014} proposed an improved procedure that avoids the complexification step altogether. Instead, it incorporates additional physical arguments and symmetry properties to derive the rotating metric in a more robust manner. In this work, we follow this improved prescription to construct an effective rotating BH geometry starting from the covariant spherically symmetric LQBH metric given in Eq.~\eqref{spherically symmetric black hole metric}.

The construction begins with a general static metric:
	\begin{equation}\label{metric non-rotating}
		ds^2=-G(r)dt^2+\frac{dr^2}{F(r)}+H(r)(d\theta^2+\sin^2\theta d\varphi^2)\,.
	\end{equation}
After introducing the advanced null coordinates $(u,r,\theta,\varphi)$ defined by
\begin{equation}
du = dt - \frac{dr}{\sqrt{F(r)G(r)}}\,,
\end{equation}
the inverse metric can be expressed in the null-tetrad form
\begin{equation}
g^{\mu\nu} =- l^\mu n^\nu - l^\nu n^\mu + m^\mu \bar{m}^\nu + m^\nu \bar{m}^\mu\,,
\end{equation}
where $\bar{m}^\mu$ is the complex conjugate of $m^\mu$. Then, the tetrad vectors chosen as
\begin{equation}\label{gmn}
\begin{aligned}
&l^\mu = \delta^\mu_r\,,\quad
n^\mu = \sqrt{\frac{F(r)}{G(r)}}\delta^\mu_u - \frac{F(r)}{2}\delta^\mu_r\,, \\
&m^\mu = \frac{1}{\sqrt{2H(r)}} \left( \delta^\mu_\theta + \frac{i}{\sin\theta}\delta^\mu_\varphi \right)\,.
\end{aligned}
\end{equation}
These vectors satisfy the standard null and orthogonality conditions:
\begin{align}
    &l^\mu l_\mu = n^\mu n_\mu = m^\mu m_\mu = \bar{m}^\mu \bar{m}_\mu = 0\,, \\
    &l^\mu n_\mu = -m^\mu \bar{m}_\mu = 1\,.
\end{align}

Next, we consider the complex coordinate transformation
\begin{equation}\label{tansf1}
r \rightarrow r + i a \cos\theta\,, \quad u \rightarrow u - i a \cos\theta\,.
\end{equation}
Instead of applying Eq.~\eqref{tansf1} directly, we perform the following transformation:
\begin{equation}\label{tansf2}
\delta^\mu_r \rightarrow \delta^\mu_r\,,\quad
\delta^\mu_u \rightarrow \delta^\mu_u\,,\quad
\delta^\mu_\theta \rightarrow \delta^\mu_\theta + i a \sin\theta (\delta^\mu_u - \delta^\mu_r)\,,\quad
\delta^\mu_\varphi \rightarrow \delta^\mu_\varphi\,.
\end{equation}
Under these transformations, the original metric functions ${G(r),F(r),H(r)}$ are extended to undetermined functions ${A(r,\theta,a),B(r,\theta,a),\Psi(r,\theta,a)}$, which must reduce to ${G(r),F(r),H(r)}$, respectively, in the non-rotating limit $a \to 0$. Correspondingly, the tetrad vectors in Eq.~\eqref{gmn} become
\begin{equation}\label{gmn2}
\begin{aligned}
&l^\mu = \delta^\mu_r\,, \quad
n^\mu = \sqrt{\frac{B}{A}}\delta^\mu_u - \frac{B}{2}\delta^\mu_r\,, \\
&m^\mu = \frac{1}{\sqrt{2\Psi}} \Bigl( \delta^\mu_\theta + i a \sin\theta (\delta^\mu_u - \delta^\mu_r) + \frac{i}{\sin\theta} \delta^\mu_\varphi \Bigr)\,.
\end{aligned}
\end{equation}
Using these tetrads, the transformed inverse metric $ g^{\mu\nu} $ can be constructed. The line element in advanced null EFCs then reads
\begin{equation}\label{metric efc}
\begin{aligned}
ds^{2} = -A \, du^{2}
&- 2 \frac{\sqrt{A}}{\sqrt{B}} \, du \, dr
- 2a \sin^{2}\theta \left( \frac{\sqrt{A}}{\sqrt{B}} - A \right) du \, d\varphi \\
&+ 2a \sin^{2}\theta \frac{\sqrt{A}}{\sqrt{B}} \, dr \, d\varphi
+ \Psi \, d\theta^{2} \\
&+ \sin^{2}\theta \left[ \Psi + a^{2} \sin^{2}\theta \left( 2\frac{\sqrt{A}}{\sqrt{B}} - A \right) \right] d\varphi^{2}\,.
\end{aligned}
\end{equation}

The final step is to bring the metric into BLCs. Following the prescription of \cite{AzregAinou2014}, we perform a coordinate transformation of the form
\begin{equation}
du = dt + \lambda(r)dr\,, \qquad d\varphi = d\phi + \chi(r)dr\,,
\end{equation}
with
\begin{equation}
\lambda(r)= -\frac{K(r)+a^{2}}{F(r)H(r)+a^{2}}\,, \qquad
\chi(r)=-\frac{a}{F(r)H(r)+a^{2}}\,,
\end{equation}
where
\begin{equation}
K(r)=\sqrt{\frac{F(r)}{G(r)}}H(r)\,.
\end{equation}
Eq.~\eqref{metric efc} reduces to the BLC form provided we choose
\begin{equation}
A(r,\theta)=\frac{\bigl[F(r)H(r)+a^{2}\cos^{2}\theta\bigr]\Psi}
{\bigl[K(r)+a^{2}\cos^{2}\theta\bigr]^{2}}\,,\qquad
B(r,\theta)=\frac{F(r)H(r)+a^{2}\cos^{2}\theta}{\Psi}\,.
\end{equation}
The resulting line element in BLCs is
	\begin{equation}
		\begin{aligned}
			d s^2& =-\frac{\left(F H+a^2 \cos ^2 \theta\right) \Psi d t^2}{\left(K+a^2 \cos ^2 \theta\right)^2}+\frac{\Psi d r^2}{F H+a^2} \\
			& -2 a \sin ^2 \theta\left[\frac{K-F H}{\left(K+a^2 \cos ^2 \theta\right)^2}\right] \Psi d t d \phi+\Psi d \theta^2 \\
			& +\Psi \sin ^2 \theta\left[1+a^2 \sin ^2 \theta \frac{2 K-F H+a^2 \cos ^2 \theta}{\left(K+a^2 \cos ^2 \theta\right)^2}\right] d \phi^2\,,
            \label{metric bl}
		\end{aligned}
	\end{equation}
By comparison with the Kerr metric, the above line element can be written in a Kerr‑like form
	\begin{equation}
		\begin{aligned}\label{kerr-like eq}
			d s^2&=-\frac{\Psi}{\rho^2}\left[\left(1-\frac{2 f}{\rho^2}\right) d t^2-\frac{\rho^2}{\Delta} d r^2\right. \\
			& \left.+\frac{4 a f \sin ^2 \theta}{\rho^2} d t d \phi-\rho^2 d \theta^2-\frac{\Sigma \sin ^2 \theta}{\rho^2} d \phi^2\right]\,,
		\end{aligned}
	\end{equation}
	where
	\begin{equation}
		\begin{aligned}
			& \rho^2 \equiv K+a^2 \cos ^2 \theta, \quad&&2 f(r) \equiv K-F H\,, \\
			& \Delta(r) \equiv F H+a^2\,, &&\Sigma \equiv\left(K+a^2\right)^2-a^2 \Delta \sin ^2 \theta\,.
		\end{aligned}
	\end{equation}
Throughout the derivation, $\Psi(r,\theta,a)$ remains an undetermined function. In the case $F(r)=G(r)$, one possible choice consistent with the procedure is $\Psi=H(r)+a^2~\cos^2\theta$ \cite{AzregAinou2014a}. 

In our model, the functions $F(r)$ and $G(r)$ are distinct. We therefore first perform a coordinate transformation to a form adapted to the improved NJ algorithm, so that the static metric \eqref{spherically symmetric black hole metric} can be rewritten as:
	\begin{equation}
		ds^2=-h(y)dt^2+\frac{1}{h(y)}dy^2+b(y)^2d\Omega^2\,,
	\end{equation}
where
\begin{equation}\label{hybb}
h(y)=f(r)=1-\frac{2M}{r}\,,\qquad
dy=\frac{dr}{\sqrt{g(r)}}=\sqrt{\frac{r}{r-r_{0}}}dr\,,\qquad
b^{2}(y)=r^{2}\,.
\end{equation}
We can follow the same steps as outlined above to obtain the rotating metric in BLCs:
	\begin{equation}
		\begin{aligned}
			d s^2& =-\frac{\left(hb^2+a^2 \cos ^2 \theta\right) \Psi d t^2}{\left(b^2+a^2 \cos ^2 \theta\right)^2}+\frac{\Psi d y^2}{hb^2+a^2} \\
			& -2 a \sin ^2 \theta\left[\frac{b^2-hb^2}{\left(b^2+a^2 \cos ^2 \theta\right)^2}\right] \Psi d t d \phi+\Psi d \theta^2 \\
			& +\Psi \sin ^2 \theta\left[1+a^2 \sin ^2 \theta \frac{2 b^2-hb^2+a^2 \cos ^2 \theta}{\left(b^2+a^2 \cos ^2 \theta\right)^2}\right] d \phi^2\,.
		\end{aligned}
	\end{equation}
For the case $F(r)=G(r)$, one has $\Psi=b^2+a^2~\cos^2\theta$. Substituting the relations in \eqref{hybb} back into the metric and simplifying, we finally obtain
	\begin{equation}
		\begin{aligned}\label{rotating metric}
			d s^2&=-\left(1-\frac{2 Mr}{\rho^2}\right) d t^2+\frac{\rho^2}{\Delta g(r)} d r^2 \\
			& -\frac{4 a Mr \sin ^2 \theta}{\rho^2} d t d \phi+\rho^2 d \theta^2+\frac{\Sigma \sin ^2 \theta}{\rho^2} d \phi^2\,,
		\end{aligned}
	\end{equation}
with the standard Kerr-type auxiliary functions
	\begin{equation}
    \begin{aligned}
        	&\Delta(r) \equiv a^2+r^2-2Mr\,,\\
            &\rho^2 \equiv r^2+a^2 \cos ^2 \theta\,,\\ &\Sigma \equiv\left(r^2+a^2\right)^2-a^2 \Delta \sin ^2 \theta\,.
    \end{aligned}
	\end{equation}
It is straightforward to verify that the metric \eqref{rotating metric} reduces to the original spherically symmetric form \eqref{spherically symmetric black hole metric} when the rotation parameter $a=0$ , and to the standard Kerr solution when the loop quantum parameter $ r_0 =0$.

\section{Scalar field dynamics and numerical method}\label{section4}

We investigate how the rotating BH described above responds to perturbations from a massless scalar field. The dynamics of the scalar field is governed by the Klein–Gordon equation
	\begin{equation}\label{kgf}
		\nabla_\mu\nabla^\mu\Phi=\frac1{\sqrt{-g}}\partial_\mu(\sqrt{-g}g^{\mu\nu}\partial_\nu\Phi)=0\,,
	\end{equation}
where $g^{\mu\nu}$ denotes the background metric. Using the ansatz
	\begin{equation}
		\Phi(t,r,\theta,\phi)=R(r)S(\theta)e^{-i\omega t+i m \phi}\,,
	\end{equation}
with $\omega$ and $ m $ being the frequency and azimuthal number of the perturbation, Eq.~\eqref{kgf} separates into a radial master equation
	\begin{equation}\label{eqr}
		\sqrt{g(r)}\frac{d}{dr}\left[\Delta\sqrt{g(r)}\frac{dR(r)}{dr}\right]+\left\{\frac{\left[(r^2+a^2)\omega-am\right]^2}{\Delta}+2am\omega-A_{lm}-a^2\omega^2\right\}R(r)=0\,,
	\end{equation}
and an angular master equation
	\begin{equation}\label{eqa}
		\frac{d}{du}\left[(1-u^2)\frac{dS(u)}{du}\right]+\left[a^2\omega^2u^2-\frac{m^2}{1-u^2}+A_{lm}\right]S(u)=0\,,
	\end{equation}
where $u=\cos\theta$ and $A_{lm}$ is the separation constant. In the subsequent numerical calculations, we set $M=1/2$ without loss of generality.

The continued fraction method (CFM), first introduced by Leaver \cite{Leaver1985}, is among the most accurate techniques for calculating QNMs and is widely employed for rotating BHs. This method \red{is} based on finding an analytical solution to Eqs.~\eqref{eqa} and~\eqref{eqr} in the form of a power series that satisfies the appropriate QNM boundary conditions. 

For the radial equation \eqref{eqr}, the boundary conditions are
	\begin{equation}
		R(r)\sim\begin{cases}
			(r-r_+)^{\frac{i(1+b)(\sqrt{1-b}m-\sqrt{1+b}m)}{2b\sqrt{1+b-2r_0}}}\,&r\rightarrow r_+\,,\\
			e^{i\omega r}r^{-1+\frac12i(2+r_0)\omega}\,&r\rightarrow \infty\,,
		\end{cases}
	\end{equation}
where $r_\pm=M\pm\sqrt{M^2-a^2}=1/2\pm\sqrt{1/4-a^2}$ are the roots of $\Delta$. By defining an auxiliary rotation parameter $b=\sqrt{1-4a^2}$, they can be further written as $r_\pm=(1\pm b)/2$. Unlike the Kerr case, the radial equation possesses singularities at $\{0,r_0,r_-,r_+,\infty\}$; consequently, a conventional power series in terms of $r\rightarrow \frac{r-r_+}{r-r_-}$ does not yield a well-defined solution over the entire domain. Following the approach adopted for the non-rotating case in this model \cite{Moreira2023}, we instead employ the mapping
	\begin{equation}
		r\rightarrow \frac{r-r_+}{r-r_0}\,.
	\end{equation}
The radial solution can then be expressed as
	\begin{equation}\label{eqrr}
            R(r)=e^{i\omega r}r_-(r-r_+)^{\xi}(r-r_0)^{-2+\frac12(2+r_0)\omega-\xi}\sum_{n=0}^{\infty}a_n(\frac{r-r_+}{r-r_0})^n\,,
	\end{equation}
where
\begin{equation}
    \xi=\frac{i(1+b)(\sqrt{1-b}m-\sqrt{1+b}m)}{2b\sqrt{1+b-2r_0}}\,,
\end{equation}
which exhibits improved convergence compared to the $r\rightarrow \frac{r-r_+}{r-r_-}$ mapping, especially for larger values of $r_0$.

The expansion coefficients $\{a_n\}_{n \in \mathbb{N}}$ satisfy a six-term recurrence relation:
	\begin{equation}
		\begin{aligned}\label{recurrence relation}
			&\alpha_0a_1+\beta_0a_0=0\,,\\
			&\alpha_1a_2+\beta_1a_1+\gamma_1a_0=0\,,\\
			&\alpha_2a_3+\beta_2a_2+\gamma_2a_1+\delta_2a_0=0\,,\\
			&\alpha_3a_4+\beta_3a_3+\gamma_3a_2+\delta_3a_1+\epsilon_3a_0=0\,,\\
			&\alpha_4a_5+\beta_4a_4+\gamma_4a_3+\delta_4a_2+\epsilon_4a_1+\sigma_4a_0=0\,,\\
			&\alpha_na_{n+1}+\beta_na_{n}+\gamma_na_{n-1}+\delta_na_{n-2}+\epsilon_na_{n-3}+\sigma_na_{n-4}=0\,.~~~(n=4,5,\cdots)\\
		\end{aligned}
	\end{equation}
Using Gaussian elimination \cite{Leaver1990}, this six-term system can be reduced to a three-term recurrence relation:
	\begin{equation}
		\begin{aligned}
			&\alpha_0'a_1+\beta_0'a_0=0\,,\\
			&\alpha_n'a_{n+1}+\beta_n'a_{n}+\gamma_n'a_{n-1}=0\,.\\
		\end{aligned}
	\end{equation}
Because the coefficients $\{\alpha_n,\beta_n,\cdots,\sigma_n\}$ are lengthy, we provide the details of the Gaussian elimination in Appendix \ref{app:A}. The series in Eq.~\eqref{eqrr} converges and the boundary condition at $r=\infty$ is satisfied if, for given parameters $a,m,$ and $A_{lm}$, the frequency $\omega$ is a root of the continued fraction equation:
	\begin{equation}\label{cfm r}
		\beta'_0-\frac{\alpha'_0 \gamma'_1}{\beta'_1-} \frac{\alpha'_1\gamma'_2}{\beta'_2-} \frac{\alpha'_2\gamma'_3}{\beta'_3-\cdots}  \equiv \beta'_0-\frac{\alpha'_0 \gamma'_1}{\beta'_1-\frac{\alpha'_1 \gamma'_2}{\beta'_2-\frac{\alpha'_2 \gamma'_3}{\beta'_3-\cdots}}}=0\,.
	\end{equation}
Similarly, the boundary conditions for the angular equation \eqref{eqa} follow from the asymptotic behavior of $S(u)$ near $u\rightarrow\pm 1$:
	\begin{equation}
		S(u)\sim\begin{cases}
			(1-u)^{m/2}\,,&u\rightarrow 1\\
			(1+u)^{m/2}\,,&u\rightarrow -1
		\end{cases}
	\end{equation}
	A solution to Eq.~\eqref{eqa} can be written as
	\begin{equation}\label{eqaa}
		S(u)=e^{a\omega u}(1+u)^{m/2}(1-u)^{m/2}\sum_{n=0}^{\infty}d_n (1+u)^n\,.
	\end{equation}
The coefficients $\{d_n\}$ obey a three-term recurrence relation:
	\begin{equation}
		\begin{aligned}
			&\alpha^{\theta}_0d_1+\beta^{\theta}_0d_0=0\,,\\
			&\alpha^{\theta}_nd_{n+1}+\beta^{\theta}_nd_{n}+\gamma^{\theta}_nd_{n-1}=0\,.~~~(n=1,2,\cdots)\\
		\end{aligned}
	\end{equation}
Since the angular equation \eqref{eqa} is identical to that of a Kerr BH, the coefficients $\{\alpha_n^\theta,\beta_n^\theta,\gamma_n^\theta\}$ take the same form as those given in Eq.~(20) of \cite{Leaver1985}. For given $a,m,$ and $\omega$, the separation constant $A_{lm}$ is determined by the root of the continued fraction
	\begin{equation}\label{cfm a}
		0=\beta_0^\theta-\frac{\alpha_0^\theta \gamma_1^\theta}{\beta_1^\theta-} \frac{\alpha_1^\theta \gamma_2^\theta}{\beta_2^\theta-} \frac{\alpha_2^\theta \gamma_3^\theta}{\beta_3^\theta-\cdots}\,.
	\end{equation}
Eqs.~\eqref{cfm r} and \eqref{cfm a} form a coupled system for the unknowns $A_{lm}$ and $\omega$. In practice, we solve them iteratively: starting with an initial guess $A_{lm}=l(l+1)$ in Eq.~\eqref{cfm r} to obtain a preliminary $\omega$, we then use this $\omega$ in Eq.~\eqref{cfm a} to update $A_{lm}$, repeating the process until both values converge.

\section{Properties of the quasinormal modes}\label{section5}
\subsection{Fundamental modes}

\begin{figure}[htbp]
		\centering
		\includegraphics[width=0.45\textwidth]{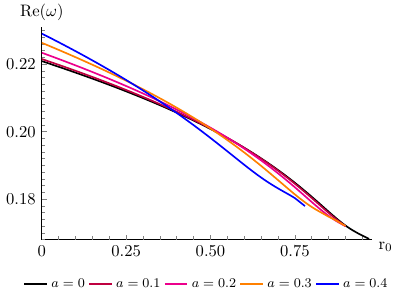}\ \hspace{0.6cm}
		\includegraphics[width=0.45\textwidth]{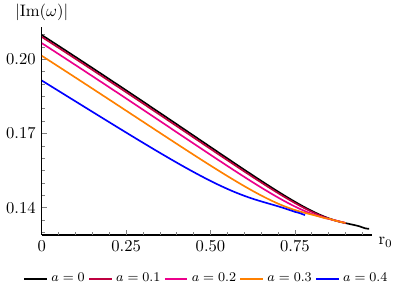}\ \\
		\caption{QNFs of the fundamental modes ($n=0$) for the scalar field perturbation with $l=0$ and $m=0$, as a function of the loop quantum parameter $r_0$. Left panel: real part $\text{Re}(\omega)$ of the QNFs; right panel: absolute value of the imaginary part $ |\text{Im}(\omega)| $. The curves are for the rotation parameter $a=0, 0.1, 0.2, 0.3, 0.4$, respectively.
        }
		\label{fm-m0-l0-n0}
	\end{figure}

\begin{figure}[htbp]
		\centering
		\includegraphics[width=0.45\textwidth]{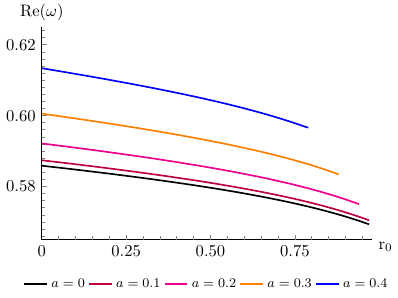}\ \hspace{0.6cm}
		\includegraphics[width=0.45\textwidth]{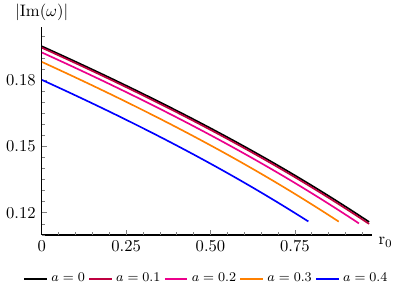}\ \\
		\caption{QNFs of the fundamental modes ($n=0$) for the scalar field perturbation with $l=1$ and $m=0$, as a function of the loop quantum parameter $r_0$. Left panel: real part $\text{Re}(\omega)$ of the QNFs; right panel: absolute value of the imaginary part $ |\text{Im}(\omega)| $. The curves are for the rotation parameter $a=0, 0.1, 0.2, 0.3, 0.4$, respectively.
        }	
		\label{fm-m0-l1-n0}
	\end{figure}

\begin{figure}[htbp]
		\centering
		\includegraphics[width=0.45\textwidth]{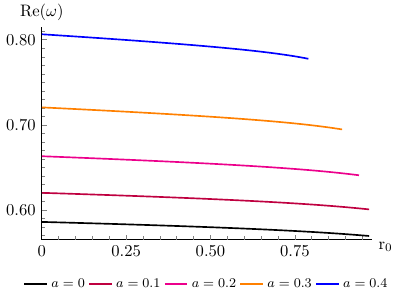}\ \hspace{0.6cm}
		\includegraphics[width=0.45\textwidth]{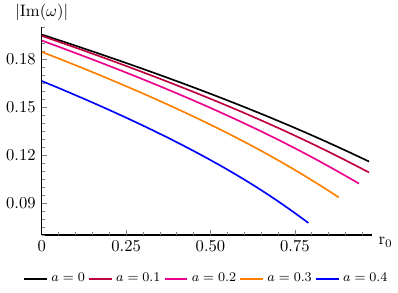}\ \\
		\caption{QNFs of the fundamental modes ($n=0$) for the scalar field perturbation with $l=1$ and $m=1$, as a function of the loop quantum parameter $r_0$. Left panel: real part $\text{Re}(\omega)$ of the QNFs; right panel: absolute value of the imaginary part $ |\text{Im}(\omega)| $. The curves are for the rotation parameter $a=0, 0.1, 0.2, 0.3, 0.4$, respectively.
        }	
		\label{fm-m1-l1-n0}
	\end{figure}

We first focus on the fundamental QNMs ($n=0$) of the massless scalar field perturbation, and systematically analyze the evolution characteristics of the QNFs with the loop quantum parameter $r_0$, rotation parameter $a$, angular quantum number $l$ and azimuthal quantum number $m$. The analysis is carried out from the simplest case of $ l=0 $, $ m=0 $, and then the angular and azimuthal quantum numbers are sequentially activated to reveal the modulation effects of different quantum numbers on the fundamental QNFs. The numerical results are presented in Figs.~\ref{fm-m0-l0-n0}, \ref{fm-m0-l1-n0}, and \ref{fm-m1-l1-n0}. Selected QNF values for various parameter combinations are also listed in Table~\ref{table1}.

For the case of $ l=0 $, $ m=0 $, the variation of the fundamental QNFs with the loop quantum parameter $r_0$ for different rotation parameters $a$ is shown in Fig.~\ref{fm-m0-l0-n0}. A universal trend can be observed that both the oscillation frequency $\text{Re}(\omega)$ and the damping rate $ |\text{Im}(\omega)| $ decrease monotonically with the increase of $r_0$, which implies that a larger loop quantum parameter $r_0$ reduces the oscillation of the scalar perturbation and retards its decay process around the rotating LQBH.

Rotation, however, leaves a distinct and non-trivial imprint on the fundamental QNFs for $ l=0 $, $ m=0 $. As shown in Fig. \ref{fm-m0-l0-n0}, as the rotation parameter $a$ increases, the oscillation frequency $\text{Re}(\omega)$ exhibits a steeper decreasing trend with the growth of $r_0$. This leads to a crossover phenomenon in the dependence of $\text{Re}(\omega)$ on $a$: for small values of $r_0$, a larger $a$ corresponds to a higher oscillation frequency $\text{Re}(\omega)$, while the opposite relationship holds for large $r_0$. In contrast, the damping rate $ |\text{Im}(\omega)| $ is generally suppressed by the increase of $a$ for all ranges of $r_0$, which means that the rotation of the BH prolongs the lifetime of the scalar perturbation. Notably, as the loop quantum parameter $r_0$ approaches its upper limit ($ r_0\rightarrow1 $), both $\text{Re}(\omega)$ and $ |\text{Im}(\omega)| $ converge to a set of common values, independent of the specific choice of the rotation parameter $a$. This convergence behavior reflects the dominant role of quantum gravitational effects over rotational effects when the LQG correction is sufficiently strong.
	
Subsequently, we turn to the case with the angular quantum number activated ($ l=1 $,$ m=0 $), and the corresponding QNF evolution is presented in Fig.~\ref{fm-m0-l1-n0}. For $ l=1 $,$ m=0 $, the monotonic decrease of $\text{Re}(\omega)$ and $ |\text{Im}(\omega)| $ with the increase of $r_0$ is still preserved, which is a robust feature of the fundamental modes in the rotating LQBH spacetime. However, the introduction of the angular quantum number $l=1$ eliminates the key characteristics observed in the $l=0,m=0$ case: the crossover phenomenon in the dependence of $\text{Re}(\omega)$ on $a$ and the convergence behavior of $\text{Re}(\omega)$ and $ |\text{Im}(\omega)| $ at the upper limit of $r_0$ disappear completely. Instead, the modulation effect of $a$ on the fundamental QNFs presents a regular trend for $ l=1 $,$ m=0 $: an increase in $a$ leads to a rise in $\text{Re}(\omega)$ and a reduction in $ |\text{Im}(\omega)| $ for a fixed $r_0$.

\begin{table}[htbp]
    \centering
    \fontsize{11}{11}\selectfont
    \begin{tabular}{|c|c|c|c|c|} \hline
        $ a $ &$ r_0 $&  $ l=0,m=0 $  & $ l=1,m=0 $& $ l=1,m=1 $ \\ 
        \hline
        0 & 0  & 0.22091 - 0.20979i &  0.58587 - 0.19532i & $\times$ \\
        & 1/10  & 0.21750 - 0.20094i& 0.58478 - 0.18835i & $\times$ \\
        & 1/4  & 0.21196 - 0.18745i  & 0.58303 - 0.17753i & $\times$ \\
        & 1/2 & 0.20080 - 0.16453i & 0.57965 - 0.15827i & $\times$ \\
        & 3/4 & 0.18486 - 0.14242i & 0.57522 - 0.13694i & $\times$ \\
        & 9/10 & 0.17219 - 0.13378i & 0.57152 - 0.12283i & $\times$ \\
        \hline
        0.1 & 0  & 0.22154 - 0.20903i & 0.58741 - 0.19466i &0.62017 - 0.19460i \\
        & 1/10  & 0.21805 - 0.20016i  & 0.58629 - 0.18769i& 0.61880 - 0.18713i\\
        & 1/4 &  0.21237 - 0.18667i  & 0.58449 - 0.17684i& 0.61673 - 0.17567i\\		
        & 1/2 & 0.20093 - 0.16377i & 0.58104 - 0.15752i& 0.61279 - 0.15514i\\
        & 3/4 & 0.18458 - 0.14190i & 0.57650 - 0.13612i& 0.60764 - 0.13213i\\
        & 9/10 & 0.17204 - 0.13386i &0.57271 - 0.12196i& 0.60326 - 0.11669i\\
        \hline
        0.2 & 0  & 0.22340 - 0.20651i & 0.59216 - 0.19253i & 0.66313 - 0.19158i\\
        & 1/10 & 0.21965 - 0.19764i  & 0.59095 - 0.18551i  & 0.66157 - 0.18377i\\
        & 1/4 & 0.21353 - 0.18415i & 0.58902 - 0.17460i& 0.65908 - 0.17155i\\
        & 1/2  & 0.20118 - 0.16139i  & 0.58532 - 0.15514i  & 0.65435 - 0.14944i\\
        & 3/4 & 0.18353 - 0.14049i & 0.58045 - 0.13352i& 0.64814 - 0.12424i\\
        & 9/10 & 0.17198 - 0.13382i & 0.57637 - 0.11921i& 0.64265 - 0.10699i\\
        \hline
        0.3 & 0  & 0.22634 - 0.20140i  & 0.60057 - 0.18826i & 0.72057 - 0.18449i\\
        & 1/10 & 0.22208 - 0.19258i  & 0.59920 - 0.18120i  & 0.71860 - 0.17615i\\
        & 1/4 & 0.21510 - 0.17922i & 0.59703 - 0.17021i& 0.71548 - 0.16301i\\	
        & 1/2  & 0.20093 - 0.15704i  & 0.59287 - 0.15053i & 0.70959 - 0.13894i\\		
        & 3/4  & 0.18142 - 0.13908i  & 0.58738 - 0.12858i & 0.70168 - 0.11066i\\
        & 8/10 & 0.17770 - 0.13723i & 0.58599 - 0.12385i& 0.69952 - 0.10428i\\
        \hline
        0.4 & 0  & 0.22907 - 0.19140i  & 0.61339 - 0.18014i  & 0.80655 - 0.16627i\\
        & 1/10  & 0.22387 - 0.18297i  & 0.61179 - 0.17309i  & 0.80393 - 0.15742i\\
        & 1/4  & 0.21532 - 0.17038i  & 0.60924 - 0.16208i & 0.79981 - 0.14334i\\
        & 1/2  & 0.19813 - 0.15084i  & 0.60438 - 0.14224i  & 0.79201 - 0.11683i\\
        & 3/4  & 0.18032 - 0.13833i  & 0.59787 - 0.11993i & 0.78059 - 0.08370i\\
        \hline
    \end{tabular}
    \caption{QNMs for the scalar field perturbations computed using the CFM. The values are listed for different combinations of the parameters $a$, $r_0$, $l$, and $m$.}\label{table1}
\end{table}

Finally, we investigate the effect of activating the azimuthal quantum number $m$ on the fundamental modes, with the typical case of $l=1, m=1$ shown in Fig.~\ref{fm-m1-l1-n0}. The activation of $m$ brings two notable changes to the fundamental QNFs. First, the oscillation frequency $\text{Re}(\omega)$ for $l=1, m=1$ (Fig.~\ref{fm-m1-l1-n0}) is significantly higher than that for $l=1, m=0$ (Fig.~\ref{fm-m0-l1-n0}) for the same $r_0$ and $a$. This enhancement effect of $m$ on $\text{Re}(\omega)$ becomes more pronounced as the rotation parameter $a$ increases, which is clearly evidenced by the gradually expanding interval between the $\text{Re}(\omega)$ curves of the two cases with the increase of $a$: the curve spacing is narrow at small $a$ values, while it widens distinctly for larger $a$ values across the entire range of $r_0$ (Fig.~\ref{fm-m0-l1-n0} and Fig.~\ref{fm-m1-l1-n0}, also see Table~\ref{table1}). Second, the azimuthal quantum number $m$ accelerates the decreasing rate of the damping rate $ |\text{Im}(\omega)| $ with $r_0$, and this acceleration effect is particularly obvious at high values of $a$. In summary, the introduction of the azimuthal quantum number $m$ further strengthens the oscillation of the scalar perturbation and slows down its decay, which is a synergistic effect of $m$ and the rotation parameter $a$ on the fundamental modes of the rotating LQBH.

\subsection{Overtones}

We further investigate the properties of the overtones ($n>0$) of the massless scalar field perturbation around the rotating LQBH, and focus on the evolution of overtone QNFs with the loop quantum parameter $r_0$, rotation parameter $a$, angular quantum number $l$ and azimuthal quantum number $m$. We also verify whether the rotating LQBH constructed via the NJ algorithm retains the typical overtone characteristics of the spherically symmetric LQBH, and analyze the modulation effect of rotation on the quantum gravity-induced overtone behaviors. The analysis is still carried out from the simplest case of $l=0,m=0$, and then the angular and azimuthal quantum numbers are sequentially activated to reveal the different regulation effects of quantum numbers on the overtones.

	\begin{figure}[htbp]
		\centering
		\includegraphics[width=0.45\textwidth]{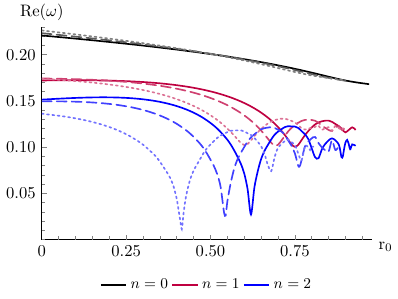}\ \hspace{0.6cm}
		\includegraphics[width=0.45\textwidth]{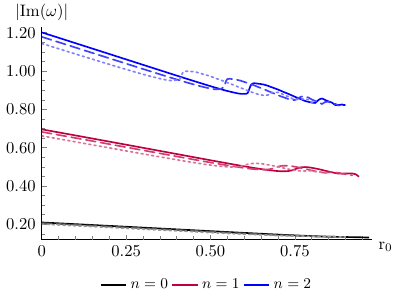}\ \\
		\caption{QNFs of scalar field perturbations ($l=0$, $m=0$) for the fundamental and overtone modes ($n=0,1,2$) as a function of the quantum parameter $r_0$. Left panel: real part $\text{Re}(\omega)$; right panel: absolute value of the imaginary part $ |\text{Im}(\omega)| $. To distinguish the rotation parameter $a$, the values $a = 0, 0.2, 0.3$ are represented by solid lines, long-dashed, and short-dashed, respectively, while color indicates the overtone order $n$.}
		\label{ot-m0-l0}
	\end{figure}
For the case of $l=0,m=0$, the variation of overtone QNFs with the loop quantum parameter $r_0$ for different rotation parameters $a$ is shown in Fig.~\ref{ot-m0-l0}. Previous studies on the spherically symmetric LQBH have revealed that the overtone QNFs exhibit obvious overtone outbursts with the increase of the principal quantum number $n$ \cite{Fu2024}. It is noteworthy that this typical quantum gravity characteristic is still preserved in the rotating LQBH constructed by the NJ algorithm, which indicates that although the NJ algorithm is applied beyond the framework of GR to construct the rotating BH solution, the model can still reflect the core quantum gravitational properties of LQBHs to a certain extent. The rotation parameter $a$ imposes a distinct regulatory effect on the overtone outburst phenomenon: for both the oscillation frequency $\text{Re}(\omega)$ and the damping rate $ |\text{Im}(\omega)| $, the introduction of $a$ causes the overtone outburst to occur in advance, i.e., the outburst phenomenon starts at a smaller value of $r_0$. In addition, for $ |\text{Im}(\omega)| $ before the occurrence of the outburst, a larger rotation parameter $a$ corresponds to a smaller value of $ |\text{Im}(\omega)| $ at the same $r_0$, which is consistent with the modulation behavior of $a$ on the damping rate of the fundamental modes. After the outburst, the behavior is consistent with previous observations in spherically symmetric LQBHs \cite{Fu2024}. As $r_0$ further increases following the outburst, the violent oscillations of the overtone QNFs gradually weaken. Both  $\text{Re}(\omega)$ and $ |\text{Im}(\omega)| $ show a convergence trend similar to that of the fundamental modes. This behavior highlights the dominant role of strong LQG corrections in governing the evolution of the overtone QNFs.

	\begin{figure}[htbp]
		\centering
		\includegraphics[width=0.45\textwidth]{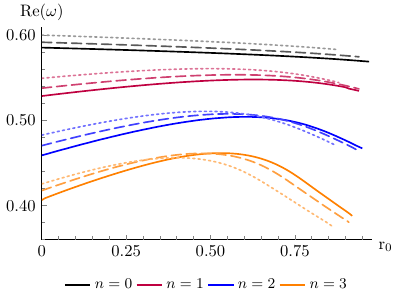}\ \hspace{0.6cm}
		\includegraphics[width=0.45\textwidth]{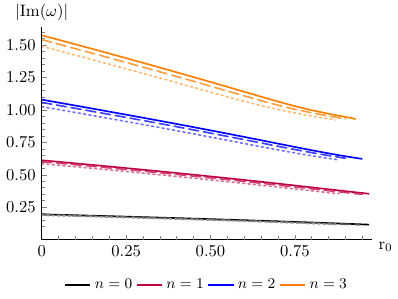}\ \\
		\caption{QNFs of scalar field perturbations ($l=1$, $m=0$) for the fundamental and overtone modes ($n=0,1,2,3$) as a function of the quantum parameter $r_0$. Left panel: real part $\text{Re}(\omega)$; right panel: absolute value of the imaginary part $ |\text{Im}(\omega)| $. To distinguish the rotation parameter $a$, the values $a = 0, 0.2, 0.3$ are represented by solid lines, long-dashed, and short-dashed, respectively, while color indicates the overtone order $n$.}
		\label{ot-m0-l1}
	\end{figure}

Subsequently, we analyze the overtone characteristics with the angular quantum number activated ($l=1,m=0$), and the corresponding QNF evolution is presented in Fig.~\ref{ot-m0-l1}. The introduction of $l=1$ leads to a key change in the overtone behavior of $\text{Re}(\omega)$: the obvious overtone outburst phenomenon observed in the $l=0, m=0$ case is significantly suppressed, and the $\text{Re}(\omega)$ of the overtones instead exhibits a non-monotonic evolution trend with the increase of $r_0$. This suppression effect of the angular quantum number $l$ on the LQG-induced overtone outburst is consistent with its modulation on the fundamental mode characteristics, reflecting the robust regulatory role of $l$ in the QNF evolution of the rotating LQBH. For the damping rate $|\text{Im}(\omega)|$ of the overtones, its evolution law is highly consistent with that of the fundamental modes: $|\text{Im}(\omega)|$ decreases monotonically with the increase of $r_0$ for all overtone numbers $n$, without obvious outburst or non-monotonic behavior, which indicates that the angular quantum number $l$ has a more significant regulatory effect on the oscillation frequency of the overtones than on the damping rate.

	\begin{figure}[htbp]
		\centering
		\includegraphics[width=0.45\textwidth]{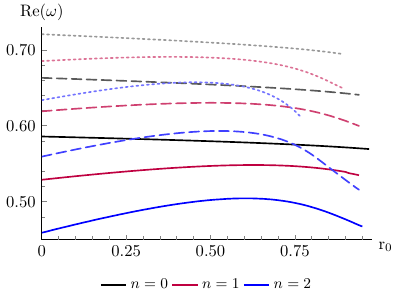}\ \hspace{0.6cm}
		\includegraphics[width=0.45\textwidth]{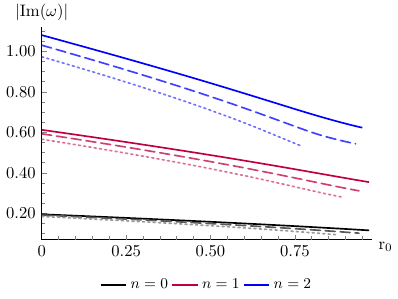}\ \\
		\caption{QNFs of scalar field perturbations ($l=1$, $m=1$) for the fundamental and overtone modes ($n=0,1,2$) as a function of the quantum parameter $r_0$. Left panel: real part $\text{Re}(\omega)$; right panel: absolute value of the imaginary part $ |\text{Im}(\omega)| $. To distinguish the rotation parameter $a$, the values $a = 0, 0.2, 0.3$ are represented by solid lines, long-dashed, and short-dashed, respectively, while color indicates the overtone order $n$.}
		\label{ot-m1-l1}
	\end{figure}

Finally, we explore the effect of activating the azimuthal quantum number m on the overtones, with the typical case of $l=0, m=0$ shown in Fig.~\ref{ot-m1-l1}. For the oscillation frequency $\text{Re}(\omega)$ of the overtones, the activation of $m=1$ basically retains the non-monotonic evolution characteristic induced by $l=1$, and the rotation parameter $a$ further amplifies its enhancement effect on $\text{Re}(\omega)$ in the overtone regime. This enhancement effect is more pronounced with the increase of the overtone number n, leading to a unique phenomenon in the $\text{Re}(\omega)$ spectrum: the oscillation frequency of the high-order overtone with a larger rotation parameter is even higher than that of the low-order overtone with a smaller rotation parameter, e.g., $\text{Re}(\omega)_{a=0.3, n=2}>\text{Re}(\omega)_{a=0.2, n=1}$. For the damping rate $|\text{Im}(\omega)|$ of the overtones, its evolution law is consistent with that of the fundamental modes under the modulation of $m$: the introduction of the azimuthal quantum number $m$ accelerates the decreasing rate of $|\text{Im}(\omega)|$ with the increase of $r_0$, and this acceleration effect is more significant in the overtone regime than in the fundamental mode regime. This phenomenon fully reflects the synergistic modulation effect of the azimuthal quantum number $m$ and the rotation parameter $a$ on the damping characteristics of the scalar field perturbation, and this effect is further amplified in the high-order overtone region.

In summary, the overtone evolution of the rotating LQBH is jointly regulated by LQG corrections, rotation effects and quantum numbers $l$, $m$: the LQG-induced overtone outburst is an inherent characteristic of the model and can be advanced by the rotation parameter $a$; the angular quantum number $l$ suppresses the overtone outburst and induces non-monotonic evolution of $\text{Re}(\omega)$; the azimuthal quantum number $m$ amplifies the enhancement effect of $a$ on $\text{Re}(\omega)$ in the overtone regime and further accelerates the attenuation of $|\text{Im}(\omega)|$ with $r_0$. These characteristics are closely related to the quantum gravitational properties of the LQBH and the rotational effect introduced by the NJ algorithm, and provide important observational signatures for the rotating LQBH.

\section{Conclusions and discussions}\label{section6}

In this work, we have constructed an effective rotating LQBH spacetime via an improved NJ algorithm from a covariant spherically symmetric LQBH model. The effective rotating LQBH is characterized by the loop quantum parameter $r_0$ and rotation parameter $a$. In this effective rotating LQBH spacetime, we systematically investigate the QNM properties of a massless probe scalar field. Our primary objective has been to elucidate the dependence of the QNFs on the quantum parameter $r_0$ and to disentangle the modulating roles played by the rotation parameter $a$, the angular quantum number $l$, and the azimuthal quantum number $m$ for both fundamental modes ($n=0$) and overtones ($n>0$). 

For the fundamental QNMs of the scalar field perturbation, our analysis reveal distinct evolutionary characteristics dictated by the quantum numbers $l$ and $m$. In the case of $l=0,m=0$, the rotation parameter $a$ modulates the change rate of the real part $\text{Re}(\omega)$ of QNFs with $r_0$, leading to a crossover phenomenon in the dependence of $\text{Re}(\omega)$ on $a$ across different regimes of $r_0$. Notably, as $r_0$ approaches its upper bound, both $\text{Re}(\omega)$ and the damping rate $|\text{Im}(\omega)|$ converge toward common asymptotic values independent of $a$. This convergence highlights the dominance of strong LQG corrections over rotational effects in the deep quantum regime. When the angular quantum number is activated ($l=1,m=0$), the aforementioned crossover and convergence features are suppressed entirely. In this scenario, the influence of rotation simplifies to a monotonic trend: increasing $a$ raises $\text{Re}(\omega)$ while reducing $|\text{Im}(\omega)|$ at a fixed value of $r_0$. The subsequent inclusion of a nonzero azimuthal quantum number ($m \neq 0$) further amplifies this rotational modulation. Specifically, $\text{Re}(\omega)$ is significantly enhanced relative to the $m=0$ case, and the rate of decrease of $|\text{Im}(\omega)|$ with respect to $r_0$ is accelerated — an effect that becomes increasingly pronounced at higher spin parameters $a$. A universal feature of the fundamental modes across all ($l$, $m$) combinations is the monotonic decrease of both $\text{Re}(\omega)$ and $|\text{Im}(\omega)|$ with increasing $r_0$. This trend indicates that stronger LQG corrections slow down the scalar field's oscillations and prolong its decay timescale in the vicinity of a rotating LQBH.

Turning to the overtone modes, we find that the effective rotating LQBH constructed via the NJ algorithm preserves the hallmark quantum gravitational signatures of its spherically symmetric counterpart. These include the distinctive overtone outburst and the nonmonotonic evolution of QNFs as a function of $r_0$. This finding substantiates the claim that, despite the extension of the NJ algorithm beyond the strict domain of GR, the resulting effective metric successfully captures essential features of LQG-induced corrections. The rotation parameter $a$ exerts a notable regulatory influence on the overtone behavior, shifting the onset of both the overtone outburst and the nonmonotonic evolution to smaller values of $r_0$ for both $\text{Re}(\omega)$ and $|\text{Im}(\omega)|$. The activation of the angular quantum number ($l=1$) tends to suppress the pronounced overtone outburst observed in the $l=0$ case, yielding a smoother non-monotonic trend in $\text{Re}(\omega)$. Meanwhile, the azimuthal quantum number $m=1$ further amplifies the enhancement effect of $a$ on $\text{Re}(\omega)$ in the overtone regime. This synergy leads to a novel spectral feature wherein the $\text{Re}(\omega)$ of a higher-order overtone with a larger spin $a$ can exceed that of a lower-order overtone with a smaller spin. 
Furthermore, the introduction of $m$ accelerates the decay of $|\text{Im}(\omega)|$ with $r_0$ for the overtones, an effect that is more pronounced than in the fundamental sector. This observation underscores the combined modulation of $m$ and $a$ on the damping characteristics of higher-order overtones.

It is important to emphasize that the direct derivation of a rotating LQBH solution from the holonomy-corrected effective equations of LQG remains an open and technically challenging problem, largely due to the complexities associated with real-valued Ashtekar–Barbero variables in axisymmetric spacetimes. The improved NJ algorithm adopted in this work provides a tractable alternative avenue for constructing an effective rotating LQBH. Our results confirm that this approach is capable of preserving the salient quantum gravitational features of LQBHs when extending to rotating configurations. Nevertheless, the future development of independent methods for generating rotating LQBH solutions will be essential for cross-validating the applicability of the NJ algorithm within the LQG framework and for deepening our understanding of the physical nature of rotating quantum BHs.

Moreover, the physical mechanism responsible for the rotation-induced advance of overtone outbursts and nonmonotonic behavior warrants further investigation. This mechanism may be intimately linked to the interplay between rotational frame-dragging effects and the quantum geometric corrections inherent to the LQBH spacetime. Looking ahead, extending this analysis to perturbations of massive scalar fields, electromagnetic fields, and Dirac fields will be valuable for assessing the universality of the observed QNM characteristics across different spin fields. Ultimately, combining the theoretical QNF spectra with future GW observational data will provide crucial phenomenological signatures for probing quantum gravity effects through BH spectroscopy. Such efforts promise to open new observational windows for testing LQG with astrophysical observations.

In summary, this work systematically unveils the QNM properties of a probe scalar field in the spacetime of an effective rotating LQBH and clarifies the multiparameter modulation laws governing the QNF spectrum. Our findings not only enrich the theoretical understanding of the interplay between quantum gravitational corrections and rotational dynamics in LQBHs but also provide a solid theoretical foundation for the astrophysical detection of rotating quantum BHs.

\section*{Acknowledgments}

This work is supported by the Natural Science Foundation of China under Grant Nos. 12375055, 12505085, the China Postdoctoral Science Foundation (No. 2025T180931), and the Jiangsu Funding Program for Excellent Postdoctoral Talent (No. 2025ZB705).

\appendix
\section{Gaussian elimination}\label{app:A}

In the CFM for computing QNFs, the radial master equation often leads to a multi-term recurrence relation for the series coefficients. To apply the standard CFM algorithm, it is necessary to reduce this multi-term recurrence to a three-term form. This appendix outlines the systematic Gaussian elimination procedure used for this reduction, first in general form and then applied explicitly to the six-term recurrence encountered in our analysis.

\subsection*{General reduction step}

Given an $i$-term recurrence relations with coefficients $C_{1,n}^{(i)},C_{2,n}^{(i)},\ldots,C_{i,n}^{(i)}$, it can be systematically reduced to an $(i-1)$-term recurrence relations $ C_{1,n}^{(i-1)},C_{2,n}^{(i-1)},\ldots,C_{i-1,n}^{(i-1)} $ by applying the following elimination step:
\begin{equation}
	\begin{aligned}
		C_{j,n}^{i-1}=C_{j,n}^{i}-\frac{C_{j,n}^{i}C_{j-1,n-1}^{i-1}}{C_{i-1,n-1}^{n-1}}
	\end{aligned}\,,
\end{equation}
where $j = 1,\dots,i-1$. The above transformation is valid for $n\ge i-2$. For $n\le i-3$, the coefficients remain unchanged: $ C_{j,n}^{(i-1)}=C_{j,n}^{(i)} $. The procedure is repeated iteratively until a three-term recurrence is obtained.

\subsection*{Example: from four-term to three-term}

For a four-term recurrence $ \{\alpha_n^{(4)},\beta_n^{(4)},\gamma_n^{(4)},\delta_n^{(4)}\} $, one elimination step yields the three-term coefficients $ \{\alpha_n^{(3)},\beta_n^{(3)},\gamma_n^{(3)}\} $:
\begin{subequations}\label{A2}
	\begin{align}
		C_n^{(3)}&=C_n^{(4)}\,,\quad\delta_n^{(4)}=0\,,\qquad &&n=0,1\,,\quad C=\alpha,\beta,\gamma\,,\\
		\alpha_n^{(3)}&=\alpha_n^{(4)}\,,&&n\ge 2\,,\\
		\beta_n^{(3)}&=\beta_n^{(4)}-\frac{\delta_n^{(4)}\alpha_{n-1}^{(3)}}{\gamma_n^{(3)}}\,,&&n\ge 2\,,\\
		\gamma_n^{(3)}&=\gamma_n^{(4)}-\frac{\delta_n^{(4)}\beta_{n-1}^{(3)}}{\gamma_n^{(3)}}\,,&&n\ge 2\,.
	\end{align}
\end{subequations}

\subsection*{Reduction of the six-term recurrence in this work}

Starting from the six-term recurrence in Eq.~\eqref{recurrence relation} of the main text, successive application of the elimination steps leads to the following three-term form
\begin{equation}
	\alpha_n'a_{n+1}+\beta_n'a_{n}+\gamma_n'a_{n-1}=0.
\end{equation}
with coefficients given explicitly by:
\begin{subequations}\label{A4}
	\begin{align}
		\alpha_n' &= \alpha_n\,, &&  \\
		\beta_n' &= \beta_n\,,&& n = 0,1\,, \\
		\beta_n' &= \beta_n - \frac{\delta_n \alpha_{n-1}'}{\gamma_{n-1}'}\,, && n = 2\,, \\
		\beta_n' &= \beta_n - \left( \delta_n - \frac{\epsilon_n \beta_{n-2}}{\gamma_{n-2}'} \right) \frac{\alpha_{n-1}'}{\gamma_{n-1}'}\,,&& n =3\,,  \\
		\beta_n' &= \beta_n - \left[ \delta_n - \frac{\epsilon_n \beta_{n-2}'}{\gamma_{n-2}'} - \sigma_n \left( \frac{\alpha_{n-3}'}{\gamma_{n-3}'} - \frac{\beta_{n-3}'  \beta_{n-2}'}{\gamma_{n-3}' \gamma_{n-2}'} \right) \right] \frac{\alpha_{n-1}'}{\gamma_{n-1}'}\,,&& n \ge4\,.
        \end{align}
\end{subequations}
\begin{subequations}\label{A5}
\begin{align}
		\gamma_n' &= 0\,,&& n = 0\,, \\
		\gamma_n' &= \gamma_n\,, && n = 1\,, \\
		\gamma_n' &= \gamma_n - \frac{\delta_n \beta_{n-1}'}{\gamma_{n-1}'}\,,&& n = 2\,, \\
		\gamma_n' &= \gamma_n - \frac{\epsilon_n \alpha_{n-2}'}{\gamma_{n-2}'} - \left( \delta_n - \frac{\epsilon_n \beta_{n-2}'}{\gamma_{n-2}'} \right) \frac{\beta_{n-1}'}{\gamma_{n-1}'}\,,&& n = 3\,, \\
		\gamma_n' &= \gamma_n - \frac{\delta_n \beta_{n-1}'}{\gamma_{n-1}'} + \epsilon_n \left( \frac{\beta_{n-2}' \beta_{n-1}'}{\gamma_{n-2}' \gamma_{n-1}'} - \frac{\alpha_{n-2}'}{\gamma_{n-2}'} \right) &&  \nonumber\\
		&\quad -\sigma_n \left( \frac{\beta_{n-3}' \beta_{n-2}' \beta_{n-1}'}{\gamma_{n-3}' \gamma_{n-2}' \gamma_{n-1}'} - \frac{\alpha_{n-3}' \beta_{n-1}'}{\gamma_{n-3}' \gamma_{n-1}'} - \frac{\beta_{n-3}'  \alpha_{n-2}'}{\gamma_{n-3}' \gamma_{n-2}'} \right)\,.&& n\ge4\,.
\end{align}
\end{subequations}
This three-term recurrence is used in the continued fraction equation \eqref{cfm r} to determine the QNFs.

	\bibliographystyle{style1}
	\bibliography{qnm}

@Article{Newman1965,
  author  = {Newman, E. T. and Janis, A. I.},
  journal = {J. Math. Phys.},
  title   = {{Note on the Kerr spinning particle metric}},
  year    = {1965},
  pages   = {915--917},
  volume  = {6},
  doi     = {10.1063/1.1704350},
}

@Article{AzregAinou2014,
  author        = {Azreg-A\"\i{}nou, Mustapha},
  journal       = {Phys. Rev. D},
  title         = {{Generating rotating regular black hole solutions without complexification}},
  year          = {2014},
  number        = {6},
  pages         = {064041},
  volume        = {90},
  archiveprefix = {arXiv},
  doi           = {10.1103/PhysRevD.90.064041},
  eprint        = {1405.2569},
  primaryclass  = {gr-qc},
}

@Article{AlonsoBardaji2022,
  author        = {Alonso-Bardaji, Asier and Brizuela, David and Vera, Ra\"ul},
  journal       = {Phys. Lett. B},
  title         = {{An effective model for the quantum Schwarzschild black hole}},
  year          = {2022},
  pages         = {137075},
  volume        = {829},
  archiveprefix = {arXiv},
  doi           = {10.1016/j.physletb.2022.137075},
  eprint        = {2112.12110},
  primaryclass  = {gr-qc},
}

@Article{AlonsoBardaji2022a,
  author        = {Alonso-Bardaji, Asier and Brizuela, David and Vera, Ra\"ul},
  journal       = {Phys. Rev. D},
  title         = {{Nonsingular spherically symmetric black-hole model with holonomy corrections}},
  year          = {2022},
  number        = {2},
  pages         = {024035},
  volume        = {106},
  archiveprefix = {arXiv},
  doi           = {10.1103/PhysRevD.106.024035},
  eprint        = {2205.02098},
  primaryclass  = {gr-qc},
}

@Article{Leaver1985,
  author  = {Leaver, E. W.},
  journal = {Proc. Roy. Soc. Lond. A},
  title   = {{An Analytic representation for the quasi normal modes of Kerr black holes}},
  year    = {1985},
  pages   = {285--298},
  volume  = {402},
  doi     = {10.1098/rspa.1985.0119},
}

@Article{Moreira2023,
  author        = {Moreira, Zeus S. and Lima Junior, Haroldo C. D. and Crispino, Lu\'\i{}s C. B. and Herdeiro, Carlos A. R.},
  journal       = {Phys. Rev. D},
  title         = {{Quasinormal modes of a holonomy corrected Schwarzschild black hole}},
  year          = {2023},
  number        = {10},
  pages         = {104016},
  volume        = {107},
  archiveprefix = {arXiv},
  doi           = {10.1103/PhysRevD.107.104016},
  eprint        = {2302.14722},
  primaryclass  = {gr-qc},
}

@Article{Leaver1990,
  author  = {Leaver, Edward W.},
  journal = {Phys. Rev. D},
  title   = {{Quasinormal modes of Reissner-Nordstrom black holes}},
  year    = {1990},
  pages   = {2986--2997},
  volume  = {41},
  doi     = {10.1103/PhysRevD.41.2986},
}

@Article{Teukolsky1973,
  author  = {Teukolsky, Saul A.},
  journal = {Astrophys. J.},
  title   = {{Perturbations of a rotating black hole. 1. Fundamental equations for gravitational electromagnetic and neutrino field perturbations}},
  year    = {1973},
  pages   = {635--647},
  volume  = {185},
  doi     = {10.1086/152444},
}

@Article{Echeverria1989,
  author  = {Echeverria, F.},
  journal = {Phys. Rev. D},
  title   = {{Gravitational Wave Measurements of the Mass and Angular Momentum of a Black Hole}},
  year    = {1989},
  pages   = {3194--3203},
  volume  = {40},
  doi     = {10.1103/PhysRevD.40.3194},
}

@Article{Finn1992,
  author        = {Finn, Lee S.},
  journal       = {Phys. Rev. D},
  title         = {{Detection, measurement and gravitational radiation}},
  year          = {1992},
  pages         = {5236--5249},
  volume        = {46},
  archiveprefix = {arXiv},
  doi           = {10.1103/PhysRevD.46.5236},
  eprint        = {gr-qc/9209010},
  reportnumber  = {PRINT-93-0128 (NORTHWESTERN)},
}

@Article{Berti2006,
  author        = {Berti, Emanuele and Cardoso, Vitor and Will, Clifford M.},
  journal       = {Phys. Rev. D},
  title         = {{On gravitational-wave spectroscopy of massive black holes with the space interferometer LISA}},
  year          = {2006},
  pages         = {064030},
  volume        = {73},
  archiveprefix = {arXiv},
  doi           = {10.1103/PhysRevD.73.064030},
  eprint        = {gr-qc/0512160},
}

@Article{Berti2018,
  author        = {Berti, Emanuele and Yagi, Kent and Yang, Huan and Yunes, Nicol\'as},
  journal       = {Gen. Rel. Grav.},
  title         = {{Extreme Gravity Tests with Gravitational Waves from Compact Binary Coalescences: (II) Ringdown}},
  year          = {2018},
  number        = {5},
  pages         = {49},
  volume        = {50},
  archiveprefix = {arXiv},
  doi           = {10.1007/s10714-018-2372-6},
  eprint        = {1801.03587},
  primaryclass  = {gr-qc},
}

@Article{Fu2024,
  author        = {Fu, Guoyang and Zhang, Dan and Liu, Peng and Kuang, Xiao-Mei and Wu, Jian-Pin},
  journal       = {Phys. Rev. D},
  title         = {{Peculiar properties in quasinormal spectra from loop quantum gravity effect}},
  year          = {2024},
  number        = {2},
  pages         = {026010},
  volume        = {109},
  archiveprefix = {arXiv},
  doi           = {10.1103/PhysRevD.109.026010},
  eprint        = {2301.08421},
  primaryclass  = {gr-qc},
}

@Article{Gong2024,
  author        = {Gong, Huajie and Li, Shulan and Zhang, Dan and Fu, Guoyang and Wu, Jian-Pin},
  journal       = {Phys. Rev. D},
  title         = {{Quasinormal modes of quantum-corrected black holes}},
  year          = {2024},
  number        = {4},
  pages         = {044040},
  volume        = {110},
  archiveprefix = {arXiv},
  doi           = {10.1103/PhysRevD.110.044040},
  eprint        = {2312.17639},
  primaryclass  = {gr-qc},
}

@Article{Zhu2025,
  author        = {Zhu, Li-Gang and Fu, Guoyang and Li, Shulan and Zhang, Dan and Wu, Jian-Pin},
  journal       = {Phys. Rev. D},
  title         = {{Quasinormal modes of a charged loop quantum black hole}},
  year          = {2025},
  number        = {10},
  pages         = {104008},
  volume        = {111},
  archiveprefix = {arXiv},
  doi           = {10.1103/PhysRevD.111.104008},
  eprint        = {2410.00543},
  primaryclass  = {gr-qc},
}

@Article{Isi2019,
  author        = {Isi, Maximiliano and Giesler, Matthew and Farr, Will M. and Scheel, Mark A. and Teukolsky, Saul A.},
  journal       = {Phys. Rev. Lett.},
  title         = {{Testing the no-hair theorem with GW150914}},
  year          = {2019},
  number        = {11},
  pages         = {111102},
  volume        = {123},
  archiveprefix = {arXiv},
  doi           = {10.1103/PhysRevLett.123.111102},
  eprint        = {1905.00869},
  primaryclass  = {gr-qc},
  reportnumber  = {LIGO-P1900135},
}

@Article{Abbott2021,
  author        = {Abbott, R. and others},
  journal       = {Phys. Rev. D},
  title         = {{Tests of general relativity with binary black holes from the second LIGO-Virgo gravitational-wave transient catalog}},
  year          = {2021},
  number        = {12},
  pages         = {122002},
  volume        = {103},
  archiveprefix = {arXiv},
  collaboration = {LIGO Scientific, Virgo},
  doi           = {10.1103/PhysRevD.103.122002},
  eprint        = {2010.14529},
  primaryclass  = {gr-qc},
  reportnumber  = {LIGO-P2000091},
}

@Article{Giesler2019,
  author        = {Giesler, Matthew and Isi, Maximiliano and Scheel, Mark A. and Teukolsky, Saul},
  journal       = {Phys. Rev. X},
  title         = {{Black Hole Ringdown: The Importance of Overtones}},
  year          = {2019},
  number        = {4},
  pages         = {041060},
  volume        = {9},
  archiveprefix = {arXiv},
  doi           = {10.1103/PhysRevX.9.041060},
  eprint        = {1903.08284},
  primaryclass  = {gr-qc},
}

@Article{Konoplya2024,
  author        = {Konoplya, R. A. and Zhidenko, A.},
  journal       = {JHEAp},
  title         = {{First few overtones probe the event horizon geometry}},
  year          = {2024},
  pages         = {419--426},
  volume        = {44},
  archiveprefix = {arXiv},
  doi           = {10.1016/j.jheap.2024.10.015},
  eprint        = {2209.00679},
  primaryclass  = {gr-qc},
}

@Article{Konoplya2022,
  author        = {Konoplya, R. A. and Zinhailo, A. F. and Kunz, J. and Stuchlik, Z. and Zhidenko, A.},
  journal       = {JCAP},
  title         = {{Quasinormal ringing of regular black holes in asymptotically safe gravity: the importance of overtones}},
  year          = {2022},
  pages         = {091},
  volume        = {10},
  archiveprefix = {arXiv},
  doi           = {10.1088/1475-7516/2022/10/091},
  eprint        = {2206.14714},
  primaryclass  = {gr-qc},
}

@Article{Konoplya2023,
  author        = {Konoplya, R. A.},
  journal       = {Phys. Rev. D},
  title         = {{Quasinormal modes in higher-derivative gravity: Testing the black hole parametrization and sensitivity of overtones}},
  year          = {2023},
  number        = {6},
  pages         = {064039},
  volume        = {107},
  archiveprefix = {arXiv},
  doi           = {10.1103/PhysRevD.107.064039},
  eprint        = {2210.14506},
  primaryclass  = {gr-qc},
}

@Article{Konoplya2023a,
  author        = {Konoplya, R. A. and Stuchlik, Z. and Zhidenko, A. and Zinhailo, A. F.},
  journal       = {Phys. Rev. D},
  title         = {{Quasinormal modes of renormalization group improved Dymnikova regular black holes}},
  year          = {2023},
  number        = {10},
  pages         = {104050},
  volume        = {107},
  archiveprefix = {arXiv},
  doi           = {10.1103/PhysRevD.107.104050},
  eprint        = {2303.01987},
  primaryclass  = {gr-qc},
}

@Article{Bhagwat2018,
  author        = {Bhagwat, Swetha and Okounkova, Maria and Ballmer, Stefan W. and Brown, Duncan A. and Giesler, Matthew and Scheel, Mark A. and Teukolsky, Saul A.},
  journal       = {Phys. Rev. D},
  title         = {{On choosing the start time of binary black hole ringdowns}},
  year          = {2018},
  number        = {10},
  pages         = {104065},
  volume        = {97},
  archiveprefix = {arXiv},
  doi           = {10.1103/PhysRevD.97.104065},
  eprint        = {1711.00926},
  primaryclass  = {gr-qc},
}

@Article{Cotesta2022,
  author        = {Cotesta, Roberto and Carullo, Gregorio and Berti, Emanuele and Cardoso, Vitor},
  journal       = {Phys. Rev. Lett.},
  title         = {{Analysis of Ringdown Overtones in GW150914}},
  year          = {2022},
  number        = {11},
  pages         = {111102},
  volume        = {129},
  archiveprefix = {arXiv},
  doi           = {10.1103/PhysRevLett.129.111102},
  eprint        = {2201.00822},
  primaryclass  = {gr-qc},
}

@Article{Lan2023,
  author        = {Lan, Chen and Yang, Hao and Guo, Yang and Miao, Yan-Gang},
  journal       = {Int. J. Theor. Phys.},
  title         = {{Regular Black Holes: A Short Topic Review}},
  year          = {2023},
  number        = {9},
  pages         = {202},
  volume        = {62},
  archiveprefix = {arXiv},
  doi           = {10.1007/s10773-023-05454-1},
  eprint        = {2303.11696},
  primaryclass  = {gr-qc},
}

@Article{Perez2017,
  author        = {Perez, Alejandro},
  journal       = {Rept. Prog. Phys.},
  title         = {{Black Holes in Loop Quantum Gravity}},
  year          = {2017},
  number        = {12},
  pages         = {126901},
  volume        = {80},
  archiveprefix = {arXiv},
  doi           = {10.1088/1361-6633/aa7e14},
  eprint        = {1703.09149},
  primaryclass  = {gr-qc},
}

@InProceedings{Modesto2007,
  author        = {Modesto, Leonardo},
  booktitle     = {{17th SIGRAV Conference}},
  title         = {{Loop quantum gravity and black hole singularity}},
  year          = {2007},
  month         = {1},
  archiveprefix = {arXiv},
  eprint        = {hep-th/0701239},
}

@Article{Ashtekar2006,
  author        = {Ashtekar, Abhay and Pawlowski, Tomasz and Singh, Parampreet},
  journal       = {Phys. Rev. Lett.},
  title         = {{Quantum nature of the big bang}},
  year          = {2006},
  pages         = {141301},
  volume        = {96},
  archiveprefix = {arXiv},
  doi           = {10.1103/PhysRevLett.96.141301},
  eprint        = {gr-qc/0602086},
  reportnumber  = {IGPG-06-2-1},
}

@Article{Ashtekar2006a,
  author        = {Ashtekar, Abhay and Pawlowski, Tomasz and Singh, Parampreet},
  journal       = {Phys. Rev. D},
  title         = {{Quantum Nature of the Big Bang: Improved dynamics}},
  year          = {2006},
  pages         = {084003},
  volume        = {74},
  archiveprefix = {arXiv},
  doi           = {10.1103/PhysRevD.74.084003},
  eprint        = {gr-qc/0607039},
  reportnumber  = {IGPG-06-7-2},
}

@Article{Ashtekar2006b,
  author        = {Ashtekar, Abhay and Pawlowski, Tomasz and Singh, Parampreet},
  journal       = {Phys. Rev. D},
  title         = {{Quantum Nature of the Big Bang: An Analytical and Numerical Investigation. I.}},
  year          = {2006},
  pages         = {124038},
  volume        = {73},
  archiveprefix = {arXiv},
  doi           = {10.1103/PhysRevD.73.124038},
  eprint        = {gr-qc/0604013},
  reportnumber  = {IGPG-06-03-2},
}

@Article{Bojowald2003,
  author        = {Bojowald, Martin and Vandersloot, Kevin},
  journal       = {Phys. Rev. D},
  title         = {{Loop quantum cosmology, boundary proposals, and inflation}},
  year          = {2003},
  pages         = {124023},
  volume        = {67},
  archiveprefix = {arXiv},
  doi           = {10.1103/PhysRevD.67.124023},
  eprint        = {gr-qc/0303072},
  reportnumber  = {CGPG-03-3-4},
}

@Article{Ashtekar2021,
  author        = {Ashtekar, Abhay and Bianchi, Eugenio},
  journal       = {Rept. Prog. Phys.},
  title         = {{A short review of loop quantum gravity}},
  year          = {2021},
  number        = {4},
  pages         = {042001},
  volume        = {84},
  archiveprefix = {arXiv},
  doi           = {10.1088/1361-6633/abed91},
  eprint        = {2104.04394},
  primaryclass  = {gr-qc},
}

@Article{Bojowald2007,
  author        = {Bojowald, Martin},
  journal       = {Phys. Rev. D},
  title         = {{Dynamical coherent states and physical solutions of quantum cosmological bounces}},
  year          = {2007},
  pages         = {123512},
  volume        = {75},
  archiveprefix = {arXiv},
  doi           = {10.1103/PhysRevD.75.123512},
  eprint        = {gr-qc/0703144},
  reportnumber  = {IGPG-07-3-5, NSF-KITP-07-55},
}

@Article{Bojowald2002,
  author        = {Bojowald, Martin},
  journal       = {Phys. Rev. Lett.},
  title         = {{Inflation from quantum geometry}},
  year          = {2002},
  pages         = {261301},
  volume        = {89},
  archiveprefix = {arXiv},
  doi           = {10.1103/PhysRevLett.89.261301},
  eprint        = {gr-qc/0206054},
  reportnumber  = {CGPG-02-6-2},
}

@Article{Modesto2010,
  author        = {Modesto, Leonardo},
  journal       = {Int. J. Theor. Phys.},
  title         = {{Semiclassical loop quantum black hole}},
  year          = {2010},
  pages         = {1649--1683},
  volume        = {49},
  archiveprefix = {arXiv},
  doi           = {10.1007/s10773-010-0346-x},
  eprint        = {0811.2196},
  primaryclass  = {gr-qc},
}

@Article{Modesto2006,
  author        = {Modesto, Leonardo},
  journal       = {Class. Quant. Grav.},
  title         = {{Loop quantum black hole}},
  year          = {2006},
  pages         = {5587--5602},
  volume        = {23},
  archiveprefix = {arXiv},
  doi           = {10.1088/0264-9381/23/18/006},
  eprint        = {gr-qc/0509078},
}

@Article{Campiglia2007,
  author        = {Campiglia, Miguel and Gambini, Rodolfo and Pullin, Jorge},
  journal       = {Class. Quant. Grav.},
  title         = {{Loop quantization of spherically symmetric midi-superspaces}},
  year          = {2007},
  pages         = {3649--3672},
  volume        = {24},
  archiveprefix = {arXiv},
  doi           = {10.1088/0264-9381/24/14/007},
  eprint        = {gr-qc/0703135},
  reportnumber  = {LSU-REL-032707, NSF-KITP-07-28},
}

@Article{Gambini2020,
  author        = {Gambini, R. and Olmedo, J. and Pullin, J.},
  journal       = {Class. Quant. Grav.},
  title         = {{Spherically symmetric loop quantum gravity: analysis of improved dynamics}},
  year          = {2020},
  number        = {20},
  pages         = {205012},
  volume        = {37},
  archiveprefix = {arXiv},
  doi           = {10.1088/1361-6382/aba842},
  eprint        = {2006.01513},
  primaryclass  = {gr-qc},
}

@Article{Boehmer2007,
  author        = {Boehmer, Christian G. and Vandersloot, Kevin},
  journal       = {Phys. Rev. D},
  title         = {{Loop Quantum Dynamics of the Schwarzschild Interior}},
  year          = {2007},
  pages         = {104030},
  volume        = {76},
  archiveprefix = {arXiv},
  doi           = {10.1103/PhysRevD.76.104030},
  eprint        = {0709.2129},
  primaryclass  = {gr-qc},
}

@Article{Chiou2008,
  author        = {Chiou, Dah-Wei},
  journal       = {Phys. Rev. D},
  title         = {{Phenomenological loop quantum geometry of the Schwarzschild black hole}},
  year          = {2008},
  pages         = {064040},
  volume        = {78},
  archiveprefix = {arXiv},
  doi           = {10.1103/PhysRevD.78.064040},
  eprint        = {0807.0665},
  primaryclass  = {gr-qc},
  reportnumber  = {IGC-08-7-1},
}

@Article{Yang2023,
  author        = {Yang, Jinsong and Zhang, Cong and Ma, Yongge},
  journal       = {Eur. Phys. J. C},
  title         = {{Shadow and stability of quantum-corrected black holes}},
  year          = {2023},
  number        = {7},
  pages         = {619},
  volume        = {83},
  archiveprefix = {arXiv},
  doi           = {10.1140/epjc/s10052-023-11800-8},
  eprint        = {2211.04263},
  primaryclass  = {gr-qc},
}

@Article{Gan2024,
  author        = {Gan, Wen-Cong and Kuang, Xiao-Mei and Yang, Zhen-Hao and Gong, Yungui and Wang, Anzhong and Wang, Bin},
  journal       = {Sci. China Phys. Mech. Astron.},
  title         = {{Nonexistence of quantum black and white hole horizons in an improved dynamic approach}},
  year          = {2024},
  number        = {8},
  pages         = {280411},
  volume        = {67},
  archiveprefix = {arXiv},
  doi           = {10.1007/s11433-024-2386-3},
  eprint        = {2212.14535},
  primaryclass  = {gr-qc},
}

@Article{Bojowald2006,
  author        = {Bojowald, Martin and Swiderski, Rafal},
  journal       = {Class. Quant. Grav.},
  title         = {{Spherically symmetric quantum geometry: Hamiltonian constraint}},
  year          = {2006},
  pages         = {2129--2154},
  volume        = {23},
  archiveprefix = {arXiv},
  doi           = {10.1088/0264-9381/23/6/015},
  eprint        = {gr-qc/0511108},
  reportnumber  = {AEI-2005-171, NI05065},
}

@Book{Gambini2011,
  author    = {Gambini, Rodolfo and Pullin, Jorge},
  publisher = {OUP Oxford},
  title     = {A first course in loop quantum gravity},
  year      = {2011},
}

@Article{AzregAinou2014a,
  author        = {Azreg-A\"\i{}nou, Mustapha},
  journal       = {Eur. Phys. J. C},
  title         = {{From static to rotating to conformal static solutions: Rotating imperfect fluid wormholes with(out) electric or magnetic field}},
  year          = {2014},
  number        = {5},
  pages         = {2865},
  volume        = {74},
  archiveprefix = {arXiv},
  doi           = {10.1140/epjc/s10052-014-2865-8},
  eprint        = {1401.4292},
  primaryclass  = {gr-qc},
}

@article{Liu:2020ola,
    author = {Liu, Cheng and Zhu, Tao and Wu, Qiang and Jusufi, Kimet and Jamil, Mubasher and Azreg-A{\"\i}nou, Mustapha and Wang, Anzhong},
    title = "{Shadow and quasinormal modes of a rotating loop quantum black hole}",
    eprint = "2003.00477",
    archivePrefix = "arXiv",
    primaryClass = "gr-qc",
    doi = "10.1103/PhysRevD.101.084001",
    journal = "Phys. Rev. D",
    volume = "101",
    number = "8",
    pages = "084001",
    year = "2020",
    note = "[Erratum: Phys.Rev.D 103, 089902 (2021)]"
}

@article{Brahma:2020eos,
    author = "Brahma, Suddhasattwa and Chen, Che-Yu and Yeom, Dong-han",
    title = "{Testing Loop Quantum Gravity from Observational Consequences of Nonsingular Rotating Black Holes}",
    eprint = "2012.08785",
    archivePrefix = "arXiv",
    primaryClass = "gr-qc",
    doi = "10.1103/PhysRevLett.126.181301",
    journal = "Phys. Rev. Lett.",
    volume = "126",
    number = "18",
    pages = "181301",
    year = "2021"
}

@article{Chen:2022nix,
    author = "Chen, Che-Yu",
    title = "{On the possible spacetime structures of rotating loop quantum black holes}",
    eprint = "2207.03797",
    archivePrefix = "arXiv",
    primaryClass = "gr-qc",
    doi = "10.1142/S0219887822501766",
    journal = "Int. J. Geom. Meth. Mod. Phys.",
    volume = "19",
    number = "11",
    pages = "2250176",
    year = "2022"
}

@article{Frodden:2012en,
    author = {Frodden, Ernesto and Perez, Alejandro and Pranzetti, Daniele and R{\"o}ken, Christian},
    title = "{Modelling black holes with angular momentum in loop quantum gravity}",
    eprint = "1212.5166",
    archivePrefix = "arXiv",
    primaryClass = "gr-qc",
    doi = "10.1007/s10714-014-1828-6",
    journal = "Gen. Rel. Grav.",
    volume = "46",
    number = "12",
    pages = "1828",
    year = "2014"
}

@article{Gambini:2020fnd,
    author = "Gambini, Rodolfo and Mato, Esteban and Pullin, Jorge",
    title = "{Axisymmetric gravity in real Ashtekar variables: the quantum theory}",
    eprint = "2001.02698",
    archivePrefix = "arXiv",
    primaryClass = "gr-qc",
    reportNumber = "LSU-REL-010820",
    doi = "10.1088/1361-6382/ab7966",
    journal = "Class. Quant. Grav.",
    volume = "37",
    number = "11",
    pages = "115010",
    year = "2020"
}

@article{Bolokhov:2023bwm,
    author = "Bolokhov, S. V.",
    title = "{Long-lived quasinormal modes and overtones{\textquoteright} behavior of holonomy-corrected black holes}",
    eprint = "2311.05503",
    archivePrefix = "arXiv",
    primaryClass = "gr-qc",
    doi = "10.1103/PhysRevD.110.024010",
    journal = "Phys. Rev. D",
    volume = "110",
    number = "2",
    pages = "024010",
    year = "2024"
}

@article{Zhu:2024wic,
    author = "Zhu, Li-Gang and Fu, Guoyang and Li, Shulan and Zhang, Dan and Wu, Jian-Pin",
    title = "{Quasinormal modes of a charged loop quantum black hole}",
    eprint = "2410.00543",
    archivePrefix = "arXiv",
    primaryClass = "gr-qc",
    doi = "10.1103/PhysRevD.111.104008",
    journal = "Phys. Rev. D",
    volume = "111",
    number = "10",
    pages = "104008",
    year = "2025"
}

@article{Alonso-Bardaji:2023niu,
    author = {Alonso-Bardaji, Asier and Brizuela, David and Vera, Ra{\"u}l},
    title = "{Singularity resolution by holonomy corrections: Spherical charged black holes in cosmological backgrounds}",
    eprint = "2302.10619",
    archivePrefix = "arXiv",
    primaryClass = "gr-qc",
    doi = "10.1103/PhysRevD.107.064067",
    journal = "Phys. Rev. D",
    volume = "107",
    number = "6",
    pages = "064067",
    year = "2023"
}

\end{document}